\documentclass[prl,aps,twocolumn,epsfig,showpacs]{revtex4}
\usepackage{graphicx,epsf}
\usepackage{amsmath}
\usepackage{amssymb}
\usepackage{bm}
\usepackage{pstricks}
\usepackage{psfrag}

\newcommand{\beq}{\begin{equation}}
\newcommand{\eeq}{\end{equation}}
\newcommand{\beqn}{\begin{eqnarray}}
\newcommand{\eeqn}{\end{eqnarray}}

\newcommand{\Amath}{\mathcal{A}}

\newcommand{\bsigma}{\mbox{\boldmath $\sigma $}}

\newcommand{\lpol}{\circlearrowright}
\newcommand{\rpol}{\circlearrowleft}

\def\bei{\begin{itemize}}
\def\eei{\end{itemize}}

\begin{document}

\title{Filling-Factor-Dependent Magnetophonon Resonance in Graphene}
\author{M. O. Goerbig,$^{1}$ J.-N. Fuchs,$^{1}$ K. Kechedzhi,$^{2}$ and
Vladimir I. Fal'ko$^{2}$}
\affiliation{$^1$Laboratoire de Physique des Solides, Univ. Paris-Sud, CNRS UMR 8502,
F-91405 Orsay, France}
\affiliation{$^2$Department of Physics, Lancaster University, Lancaster, LA1 4YB, United
Kingdom}
\date{\today}

\begin{abstract}
We describe a peculiar fine structure acquired by the in-plane optical
phonon at the $\Gamma $-point in graphene when it is brought into resonance
with one of the inter-Landau-level transitions in this material. The effect
is most pronounced when this lattice mode (associated with the G-band in
graphene Raman spectrum) is in resonance with inter-Landau-level transitions
$0\Rightarrow +,1$ and $-,1\Rightarrow 0$, at a magnetic field $B_{0}\simeq
30$T. It can be used to measure the strength of the electron-phonon coupling
directly, and its filling-factor dependence can be used experimentally to
detect circularly polarized lattice vibrations.
\end{abstract}

\maketitle


In metals and semiconductors the spectra of phonons are renormalized by
their interaction with electrons. Some of the best known examples include
the Kohn anomaly \cite{kohn} in the phonon dispersion, which originates from
the excitation/de-excitation of electrons across the Fermi level upon the
propagation of a phonon through the bulk\ of a metal and a shift in the
longitudinal optical phonon frequency in heavily doped polar semiconductors
\cite{mahan}. However, despite the transparency of theoretical models the
observation of such effects is often obscured by the difficulty to change
the electron density in a material, whereas in semiconductor structures
containing two-dimensional (2D) electrons the density of which can be
varied, the influence of the latter on the phonon modes is weak due to a
negligibly small volume fraction occupied by the electron gas. In this
context, a unique opportunity arises in graphene-based field-effect
transistors \cite{geim}, where the density of carriers in an atomically thin
film (monolayer \cite{novoselov1,zhang1,zhang2} or a bilayer \cite{geim2})
can be continuously varied from 10$^{13}$cm$^{-2}$ p-type to 10$^{13}$cm$%
^{-2}$ n-type. Several Raman experiments have already been reported \cite%
{pisana,yan} where the variation of carrier density in graphene changes the
optical phonon frequency, in agreement with theoretical expectations \cite%
{ando,castroneto,lazzeri}.

When graphene is exposed to a quantizing magnetic field, its electronic
spectrum quenches into discrete Landau levels (LLs) \cite{McClure}. Then,
the optical phonon energy in graphene may coincide with the energy of one of
the inter-LL transitions, a condition known as magnetophonon resonance \cite%
{magnetophonon,Nicholas}. Recently, Ando has suggested \cite{AndoMP} that
in undoped graphene the magnetophonon resonance enhances the effect of the
electron-phonon coupling on a spectrum of the in-plane optical phonons - the
E$_{2g}$ modes attributed to the G-band in the Raman spectra in Refs. \cite%
{ferrari,gupta,graf,pisana,yan}. In this paper, we investigate a rich
structure of the anti-crossing experienced by such lattice modes when a
magnetic field makes their energy equal to the energy of one of the
valley-antisymmetric interband magnetoexcitons \cite{magnetooptics}. Most
saliently, the difference between circular polarization of various inter-LL
transitions \cite{sadowski,Abergel} makes the magnetophonon resonance
distinguishable for lattice vibrations of different circular polarization,
which makes the number of split lines in the fine structure acquired by a
phonon and the value of splitting dependent on the electronic filling
factor, $\nu $.

The in-plane optical phonons in graphene [relative displacement $\mathbf{u}%
=(u_{x},u_{y})$ of sublattices $A$ and $B$] have the energy $\omega \approx
0.2$eV at the $\Gamma $-point (in the center of the Brillouin zone). These
phonons and their coupling to electrons can be described using the
Hamiltonian \cite{ando,castroneto},
\begin{eqnarray}
H_{\mathrm{ph}} &=&\sum_{\mu ,\mathbf{q}}\omega b_{\mu ,\mathbf{q}}^{\dagger
}b_{\mu ,\mathbf{q}}+g\sqrt{2M\omega }(\sigma _{x}u_{y}-\sigma _{y}u_{x}),
\label{eq01} \\
\mathbf{u}(\mathbf{r}) &=&\sum_{\mu ,\mathbf{q}}\frac{1}{\sqrt{%
2N_{uc}M\omega }}\left( b_{\mu ,\mathbf{q}}+b_{\mu ,-\mathbf{q}}^{\dagger
}\right) \mathbf{e}_{\mu ,\mathbf{q}}e^{-i\mathbf{q}\cdot \mathbf{r}},
\notag
\end{eqnarray}%
where $b_{\mu ,\mathbf{q}}^{(\dagger )}$ are annihilation (creation)
operators of a phonon with polarisation $\mathbf{e}_{\mu ,\mathbf{q}}$, $M$
is the mass of a carbon atom, and $N_{uc}$ is the number of unit cells. Here
and below, we use units $\hbar \equiv 1$. Also, we shall utilize a double
degeneracy of the E$_{2g}$ mode at the $\Gamma $-point (at $\mathbf{q}=0$)
and describe the in-plane optical phonon in terms of a degenerate pair of
circularly polarized modes, $u_{\circlearrowleft }=(u_{x}+iu_{y})/\sqrt{2}$ and $%
u_{\circlearrowright }=u_{\circlearrowleft }^*$. The constant $g$ in Eq. (\ref{eq01})
characterizes the electron-phonon coupling \cite{footnote1}. This coupling
has the form of the only invariant linear in $\mathbf{u}$ permitted by the
symmetry group of the honeycomb crystal. It is constructed using Pauli
matrices ${\bsigma }=(\sigma _{x},\sigma _{y})$ acting in the space of
sublattice components of the Bloch functions, ${[}\phi _{\mathbf{K}_{+}A}$, $%
\phi _{\mathbf{K}_{+}B}]$ and $[\phi _{\mathbf{K}_{-}B}$, $\phi _{\mathbf{K}%
_{-}A}]$ which describe electron states in the valleys $\mathbf{K}_{\pm }$
(two opposite corners of the hexagonal Brillouin zone) and obey the
Hamiltonian, in terms of the electron charge $-e<0$ \cite{footnote2},
\begin{equation*}
H_{\mathrm{el}}=\xi v\mbox{\boldmath $\sigma $}\cdot \mathbf{p},~~\mathbf{%
\;p=}-i\mathbf{\nabla }+e\mathbf{A,\;}~~\partial _{x}A_{y}-\partial
_{y}A_{x}=B.
\end{equation*}%
Here, $\xi =\pm $ distinguishes between $\mathbf{K}_{\pm }$, and momentum $%
\mathbf{\mathbf{p}}$ is calculated with respect to the center of the
corresponding valley. This Hamiltonian represents the dominant term of the
next-neighbor tight-binding model of graphene \cite%
{wallace,Dresselhaus,AndoReview}, and the electron-phonon coupling in Eq. (%
\ref{eq01}) takes into account the change in the $A-B$ hopping elements due
to the sublattice displacement \cite{symmetry}.

In a perpendicular magnetic field, $H_{\mathrm{el}}$ determines \cite%
{McClure} a spectrum of 4-fold (spin and valley) degenerate LLs, $%
\varepsilon _{n}^{\alpha=\pm }=\alpha \sqrt{2n}v\lambda _{B}^{-1}$ in the
valence band ($\varepsilon _{n>0}^{-}$), conduction band ($\varepsilon
_{n>0}^{+}$), and at zero energy ($\varepsilon _{0}=0$, exactly at the Dirac
point in the electron spectrum), in terms of the magnetic length $\lambda
_{B}=1/\sqrt{eB} $. Such a spectrum has been confirmed by recent quantum Hall effect
measurements \cite{novoselov1,zhang1,zhang2}. In each of the two valleys,
the LL basis is given by two-component states $\sqrt{\frac{1}{2}}[\sqrt{%
1+\delta _{n,0}}\phi _{n,m},i\xi \alpha (1-\delta _{n,0})\phi _{n-1,m}]$,
where $\phi _{n,m}$ are the LL wave functions described by the quantum
numbers $n$ and $m$, the latter being related to the guiding center degree
of freedom. Here, we neglect the Zeeman effect, and simply take into account
the two-fold spin degeneracy.

\begin{figure}[tbp]
\centering
\includegraphics[width=8.0cm,angle=0]{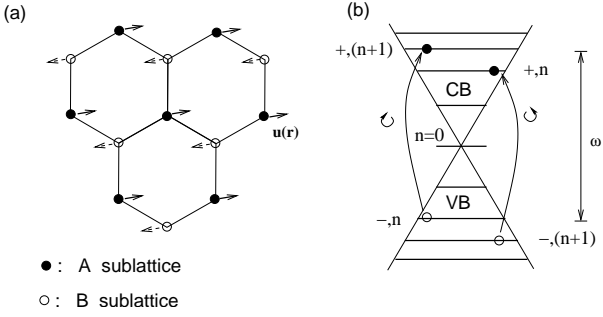}
\caption{{\protect\footnotesize {\textsl{(a)} Optical phonons are lattice
vibrations with an out-off-phase oscillation of the two sublattices. \textsl{%
(b)} Interband electron-hole excitations coupling to phonon modes with
different circular polarization. }}}
\label{fig01}
\end{figure}

Excitations of electrons between LLs can be described in terms of
magnetoexcitons (see Fig. 1). Those relevant for the magnetophonon resonance
are
\begin{eqnarray}
\psi _{\circlearrowleft }^{\dagger }(n,\xi ) &=&\frac{i\sqrt{1+\delta _{n,0}}%
}{\mathcal{N}_{n}^{\circlearrowleft }}\sum_{m}c_{+,n,m;\xi }^{\dagger
}c_{-,(n+1),m;\xi },  \notag  \label{eq02} \\
\psi _{\circlearrowright }^{\dagger }(n,\xi ) &=&\frac{i\sqrt{1+\delta _{n,0}%
}}{\mathcal{N}_{n}^{\circlearrowright }}\sum_{m}c_{+,(n+1),m;\xi }^{\dagger
}c_{-,n,m;\xi },
\end{eqnarray}%
where the index $\mathcal{A}=\circlearrowleft ,\circlearrowright $
characterizes the angular momentum of the excitation and the operators $%
c_{\alpha ,n,m;\xi }^{(\dagger )}$ annihilate (create) an electron in the
state $\alpha ,n,m$ in the valley $\mathbf{K}_{\xi }$. The normalization
factors $\mathcal{N}_{n}^{\circlearrowleft }=[(1+\delta _{n,0})N_{B}(\bar{\nu%
}_{-,(n+1)}-\bar{\nu}_{+,n})]^{1/2}$ and $\mathcal{N}_{n}^{%
\circlearrowright }=[(1+\delta _{n,0})N_{B}(\bar{\nu}_{-,n}-\bar{\nu}%
_{+,(n+1)})]^{1/2}$ are used to ensure the bosonic commutation relations of
the exciton operators, $[\psi _{\mathcal{A}}(n,\xi ),\psi _{\mathcal{A}%
^{\prime }}^{\dagger }(n^{\prime },\xi ^{\prime })]=\delta _{\mathcal{A},%
\mathcal{A}^{\prime }}\delta _{\xi ,\xi ^{\prime }}\delta _{n,n^{\prime }}$,
where $N_{B}$ is the total number of states per LL in a sample, including
the two-fold spin-degeneracy. These commutation relations are obtained
within the mean-field approximation with $\langle c_{\alpha ,n,m;\xi
}^{\dagger }c_{\alpha ^{\prime },n^{\prime },m^{\prime };\xi ^{\prime
}}\rangle =\delta _{\xi ,\xi ^{\prime }}\delta _{\alpha ,\alpha ^{\prime
}}\delta _{n,n^{\prime }}\delta _{m,m^{\prime }}(\delta _{\alpha ,-}+\delta
_{\alpha ,+}\bar{\nu}_{\alpha ,n})$, where $0\leq\bar{\nu}_{\alpha ,n}\leq 1
$ is the partial filling factor of the $n$-th LL. Similarly to
magneto-optical selection rules in graphene \cite%
{magnetooptics,sadowski,Abergel}, $\alpha,n\Rightarrow \alpha^{\prime},n\pm
1 $, $\circlearrowleft $-polarized phonons are coupled to electronic
transitions with $-,(n+1)\Rightarrow +,n$, and $\circlearrowright $%
-polarized phonons to $-,n\Rightarrow +,(n+1)$ magneto-excitons, at the same
energy $\Omega _{n}\equiv \sqrt{2}(v/\lambda _{B})(\sqrt{n}+\sqrt{n+1})$
(Fig. \ref{fig01}), which follows directly from the composition of the LL in
graphene and the form of the electron-phonon coupling in Eq. (\ref{eq01}).
In contrast to photons that couple to the valley-symmetric mode $\psi _{%
\mathcal{A},s}(n)=[\psi _{\mathcal{A}}(n,\mathbf{K}_{+})+\psi _{\mathcal{A}%
}(n,\mathbf{K}_{-})]/\sqrt{2}$, electron-phonon interaction in Eq.(\ref{eq01}%
) couples phonons to the valley-antisymmetric magnetoexciton $\psi _{%
\mathcal{A},as}(n)=[\psi _{\mathcal{A}}(n,\mathbf{K}_{+})-\psi _{\mathcal{A}%
}(n,\mathbf{K}_{-})]/\sqrt{2}$.

In terms of magnetoexcitons we can, now, rewrite the electron-phonon
Hamiltonian in a bosonized form, as
\begin{gather}
H=\sum_{\tau =s,as}\sum_{\mathcal{A},n}\Omega _{n}\psi _{\mathcal{A},\tau
}^{\dagger }(n)\psi _{\mathcal{A},\tau }(n)+\sum_{\mathcal{A}}\omega b_{%
\mathcal{A}}^{\dagger }b_{\mathcal{A}}  \label{eq03} \\
\;\;\;\;\;\;\;\;\;\;+\sum_{\mathcal{A},n}g_{\mathcal{A}}(n)\left[ b_{%
\mathcal{A}}^{\dagger }\psi _{\mathcal{A};as}(n)+b_{\mathcal{A}}\psi _{%
\mathcal{A};as}^{\dagger }(n)\right] ,  \notag \\
g_{\circlearrowleft }(n)=g\sqrt{(1+\delta _{n,0})\gamma }\sqrt{%
\bar{\nu}_{-,(n+1)}-\bar{\nu}_{+,n}},  \notag \\
g_{\circlearrowright }(n)=g\sqrt{(1+\delta _{n,0})\gamma }\sqrt{%
\bar{\nu}_{-,n}-\bar{\nu}_{+,(n+1)}},  \notag
\end{gather}%
where $g_{\mathcal{A}}$ are the effective coupling constants, with $\gamma =3%
\sqrt{3}a^{2}/2\pi \lambda _{B}^{2}$ and $a=1.4$\AA\ (distance between
neighboring carbon atoms). In the Hamiltonian (\ref{eq03}), we have omitted
electronic excitations with a higher angular momentum which do not couple to
the in-plane optical phonon modes (\textit{e.g.}, $n\Rightarrow n^{\prime}$,
with $n^{\prime}\neq n\pm 1$).
The dressed phonon operator corresponding to the Hamiltonian (\ref{eq03}) is
obtained by solving Dyson's equation. The pole of the propagator gives the
antisymmetric coupled mode
frequencies $\tilde{\omega}_{\mathcal{A}}$,
\begin{equation}
\tilde{\omega}_{\mathcal{A}}^{2}-\omega ^{2}=4\omega \left[
\sum_{n=n_{F}+1}^{N}\frac{\Omega _{n}g_{\mathcal{A}}^{2}(n)}{\tilde{\omega}_{%
\mathcal{A}}^{2}-\Omega _{n}^{2}}+ \frac{\Delta_{n_F}g_{\mathcal{A}}^2(n_F)}{
\tilde{\omega}_{\mathcal{A}}^2-\Delta_{n_F}^2} \right],  \label{eq06}
\end{equation}%
where $n_{F}$ stands for the number of the highest fully occupied LL in the
spectrum, and $\Delta_n=\sqrt{2}(v/\lambda_B)(\sqrt{n+1}-\sqrt{n})$. In Eq. (%
\ref{eq06}), the sum (extended up to the high-energy cut-off $N\sim
(\lambda_B/a)^2$ above which the electronic dispersion is no longer linear)
takes into account interband magnetoexcitons, and the last term gives a
small correction due to an intraband magnetoexciton. In the small-field
limit and large doping ($n_{F}\gg 1$), solution of Eq. (\ref{eq06})
reproduces the zero-field result \cite{ando,castroneto} if one replaces the
sum by an integral, $\sum_{n=0}^{n_{F}}\rightarrow \int_{0}^{n_{F}}dn$,
approximates $\sqrt{n}+\sqrt{n+1}\approx 2\sqrt{n}$ and $\Delta_{n_F}%
\approx0 $, and, then, linearizes Eq. (\ref{eq06}) by replacing $\tilde{%
\omega}_{\mathcal{A}}$ by $\omega $ in the denominator,
\begin{eqnarray*}
\tilde{\omega} &\simeq &\tilde{\omega}_{0}+\lambda \left[ \sqrt{2n_{F}}\frac{%
v}{\lambda _{B}}-\frac{\omega }{4}\ln \left( \frac{\omega +2\sqrt{2n_{F}}%
v/\lambda _{B}}{\omega -2\sqrt{2n_{F}}v/\lambda _{B}}\right) \right] , \\
\tilde{\omega}_{0} &\simeq &\omega +2\int_{0}^{N}dn\frac{\Omega _{n}g_{%
\mathcal{A}}^{2}(n)}{\omega ^{2}-\Omega _{n}^{2}},
\end{eqnarray*}%
where $\lambda =(2/\sqrt{3}\pi )(g/t)^{2}\simeq 3.3\times 10^{-3}$ is
the same as in
Refs. \cite{ando,AndoMP} ($t=2v/3a\sim 3$eV is the $A-B$ hopping amplitude), and $%
\tilde{\omega}_{0}$ is the renormalized phonon frequency in an undoped
graphene sheet at $B=0$. The only variation arises at high fields, $\tilde{%
\omega} _{0}\gtrsim \sqrt{2}v/\lambda _{B}$, where for $n_{F}=0$ the
linearized Eq. (\ref{eq06}) yields
\begin{equation*}
\tilde{\omega}\simeq \tilde{\omega}_{0}-\frac{\lambda _{B}\sqrt{2}}{v}\frac{%
g^{2}(0)}{(\tilde{\omega} _{0}\lambda _{B}/\sqrt{2}v)^{2}-1}.
\end{equation*}

The strongest effect of the phonon coupling to electron modes occurs when
the frequency of the former coincides with the frequency $\Omega _{n}$ of
one of the magnetoexcitons $\psi _{\mathcal{A},as}(n)$. In such a case, the
sum on the right-hand-side of the eigenvalue equation (\ref{eq06}) is
dominated by the resonance term and may be approximated by $2\omega g_{%
\mathcal{A}}^{2}(n)/\left( \tilde{\omega}_{\mathcal{A}}-\Omega _{n}\right) $%
. This results in a fine structure of mixed phonon-magnetoexciton modes, $%
\psi _{\mathcal{A},as}(n)\cos \theta +b_{\mathcal{A}}\sin \theta $ with
frequency $\tilde{\omega}_{\mathcal{A}}^{+}$ and $\psi _{\mathcal{A}%
,as}(n)\sin \theta -b_{\mathcal{A}}\cos \theta $ with frequency $\tilde{%
\omega}_{\mathcal{A}}^{-}$ [where $\cot 2\theta =(\Omega _{n}-\tilde{\omega}%
_{0})/2g_{\mathcal{A}}$], which are determined for each polarisation ($%
\mathcal{A}=\circlearrowleft ,\circlearrowright $ ) separately,
\begin{equation}
\tilde{\omega}_{\mathcal{A}}^{\pm }(n)=\tfrac{1}{2}\left( \Omega _{n}+\tilde{%
\omega}_0\right) \mp \sqrt{\tfrac{1}{4}(\Omega _{n}-\tilde{\omega}_0)^{2}+g_{%
\mathcal{A}}^{2}(n)}.  \label{eq05}
\end{equation}

\begin{figure}[tbp]
\centering
\hspace*{0.2cm}\includegraphics[width=7.5cm,angle=0]{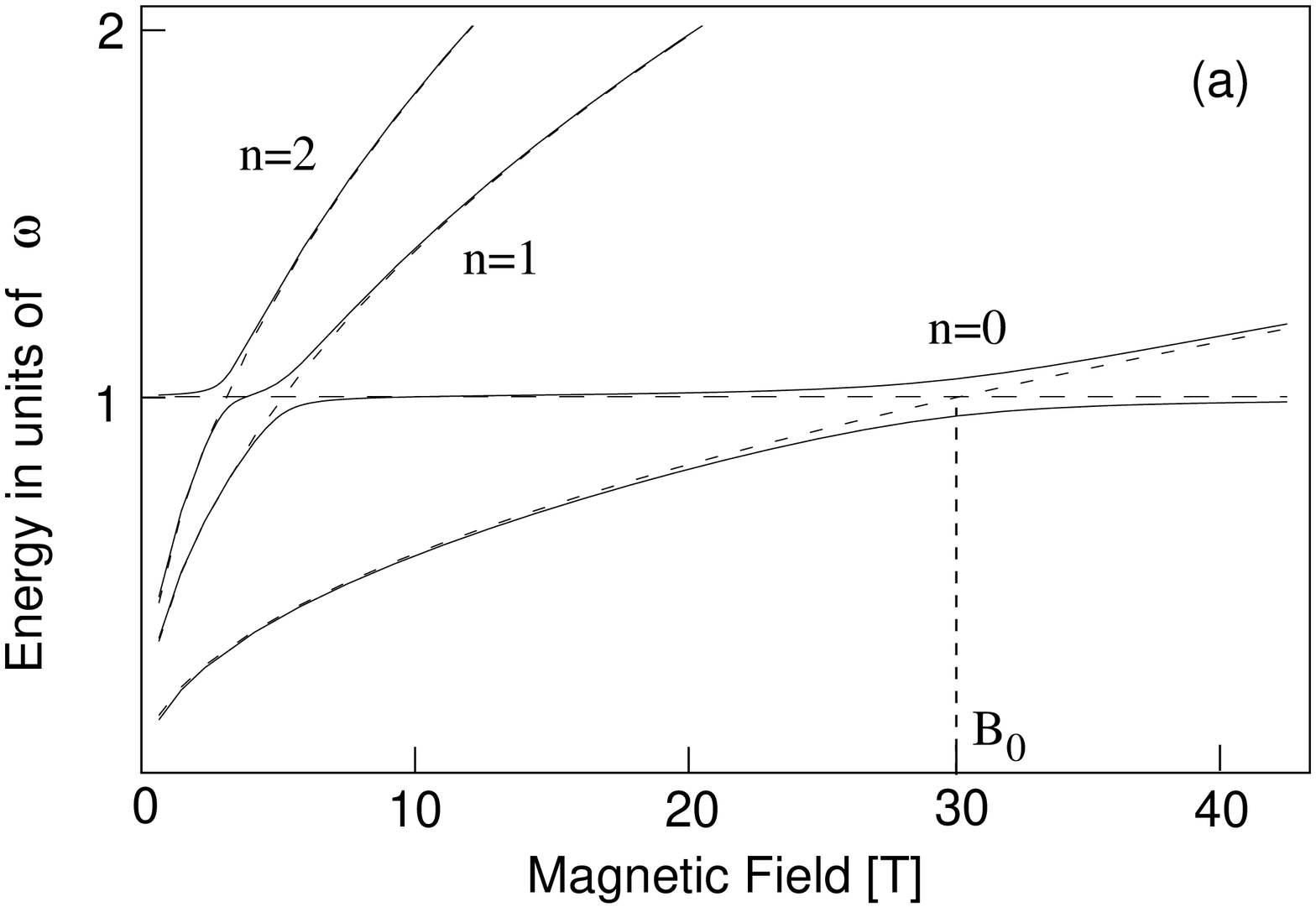} %
\includegraphics[width=6.5cm,angle=0]{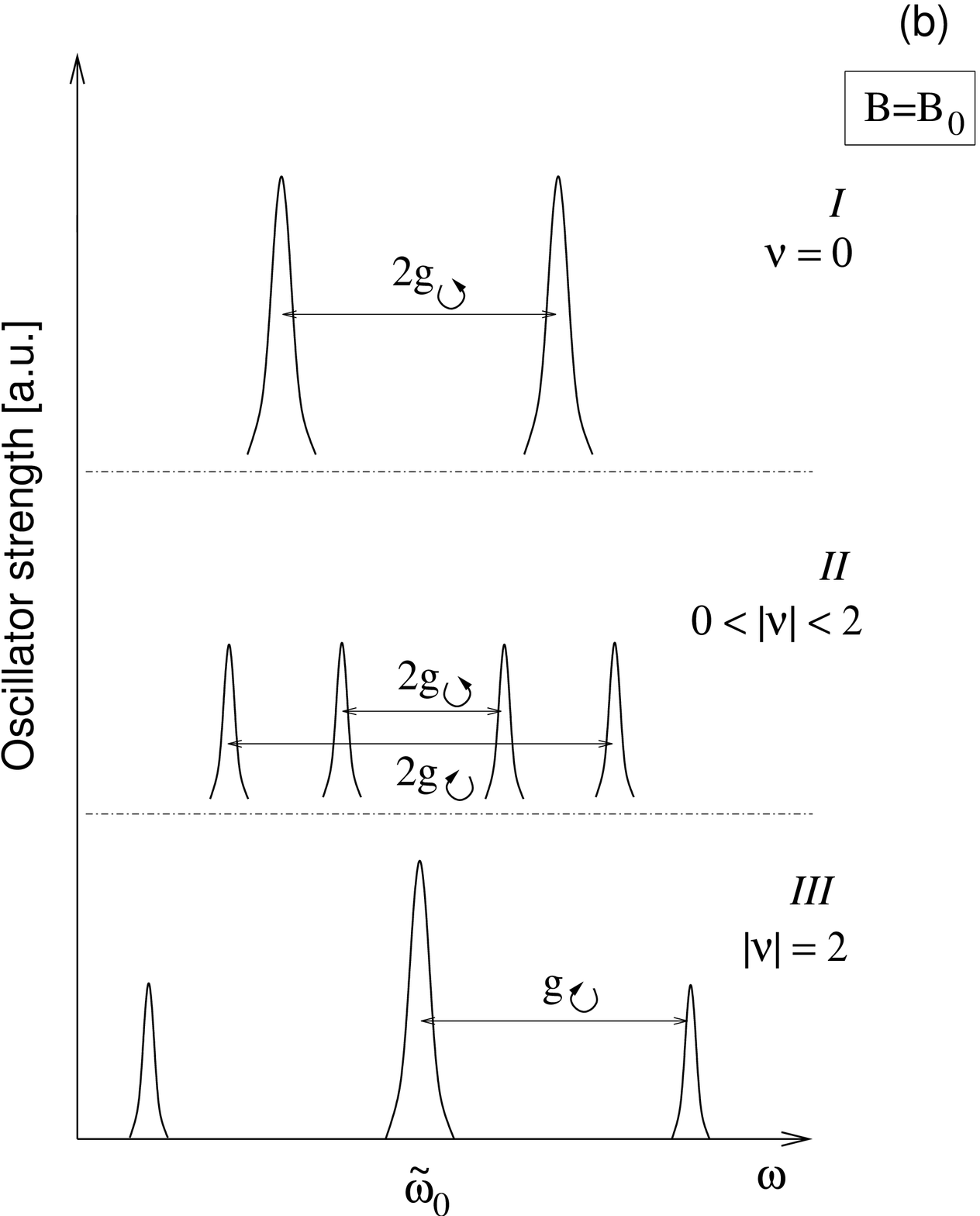} \hspace*{-0.8cm}%
\includegraphics[width=7.0cm,angle=0]{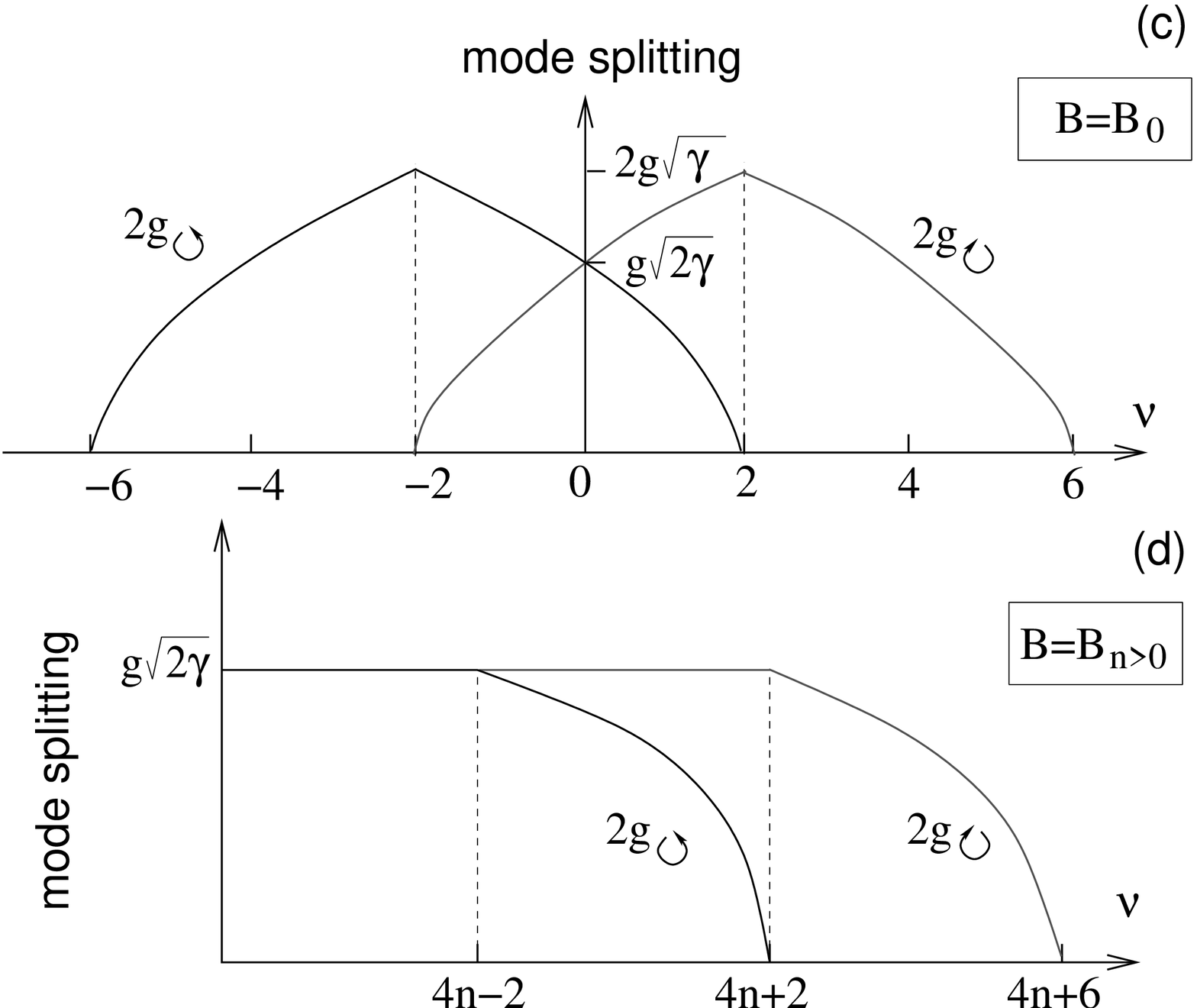}
\caption{{\protect\footnotesize {\textsl{(a)} Coupled phonon and
magneto-excitons as a function of the magnetic field. Energies are in units
of the bare phonon energy $\protect\omega$. Dashed lines indicate the
uncoupled valley-symmetric modes, with $g_{\mathcal{A}}=0$. \textsl{(b)}
Mode splitting as a function of the filling factor, as may be seen in Raman
spectroscopy, with the resonance condition $\Omega_{n=0}\approx\tilde{%
\protect\omega}_0$, for $\protect\nu=0$ in \textsl{(I)}, $0<|\protect\nu|<2$
(in \textsl{II}), and $\protect\nu=\pm 2$ (in \textsl{III}). The absolute
intensity of the modes is in arbitrary units, but the height and the width
reflect the expected relative intensities. \textsl{(c)} Mode splitting for $%
n=0$, as a function of the filling factor $\protect\nu$. \textsl{(d)} Same
as in \textsl{(c)} for $n\geq 1$.}}}
\label{fig02}
\end{figure}

A generic form of the phonon-magnetoexciton anticrossing and formation of
coupled modes, $\omega _{\mathcal{A}}^{\pm }(n)$ in undoped graphene (%
\textit{i.e.}, $\nu =0$) is illustrated in Fig. \ref{fig02}(a). Such an
anticrossing and mode mixing is simlar to that described by Ando \cite%
{AndoMP}. It can manifest itself in Raman spectroscopy: in a fine structure
acquired by the G-line (earlier attributed \cite%
{pisana,yan,ferrari,gupta,graf} to the in-plane optical phonon at the $%
\Gamma $-point, E$_{2g}$ mode) at the magneto-phonon resonance conditions.
The effect is the strongest for the resonance $\Omega _{n=0}\approx \tilde{%
\omega}_{0}$ between the phonon and magnetoexciton based upon $%
-,1\Rightarrow 0$ and $0\Rightarrow +,1$ transitions. When approaching the
resonance (by sweeping a magnetic field), the phonon line becomes
accompanied by a weak satellite moving towards it and increasing its
intensity. Exactly at the magnetophonon resonance, where both the upper mode
[$\tilde{\omega}_{\mathcal{A}}^{+}(n)$] and the lower mode [$\tilde{\omega}_{%
\mathcal{A}}^{-}(n)$] consist of an equal-weight superposition of the phonon
and the resonant exciton, with $\cos \theta =\sin \theta =1/\sqrt{2}$, the
G-band in graphene would appear as two lines. For $\Omega _{n=0}=\sqrt{2}%
v/\lambda _{B}\approx 36\sqrt{B\mathrm{[T]}}$ meV (see \cite%
{footnote2,AndoMP}) and $\tilde{\omega}_{0}\simeq 200$ meV, this resonance
occurs in an experimentally accessible field range, $B_{0}\simeq 30$ T. For
the filling factor $\nu =0$, the central LL ($n=0$) is always half-filled.
Then, coupling and, therefore, splitting of the $\circlearrowright $- and $%
\circlearrowleft $-polarized modes coincide, $g_{\circlearrowright
}=g_{\circlearrowleft }$, thus, giving rise to a pair of peaks at
the energies $\tilde{\omega}^{\pm }=\tilde{\omega}_{0}\pm
g_{\circlearrowright }$ sketched in part \textsl{I} in Fig.
\ref{fig02}(b). For the magnetic field value $B_{0}\simeq 30$ T and
$g\simeq 0.28$eV \cite{lazzeri}, we estimate this splitting as
$2g_{\mathcal{A}}\sim 16$meV ($\sim 130$cm$^{-1}$), which
largely exceeds the G-band width observed in Refs. \cite%
{pisana,yan,ferrari,gupta,graf}.

Doping of graphene changes the strength of the coupling constants $%
g_{\circlearrowright }\ $and $g_{\circlearrowleft }$, as shown in Fig. \ref%
{fig02}(c). This is because a higher (lower) occupancy of the $n=0$ LL
reduces (enhances) the oscillator strength of the $\circlearrowleft $
polarized transition due to the availability of filled and empty states in
the involved LLs, whereas the same change in the electron density has the
opposite effect on $g_{\circlearrowright }$. As a result, for an arbitrary
filling factor $-2<\nu <2$, we predict that, in the vicinity of
magnetophonon resonance, the phonon mode (and, therefore, G-band in Raman
spectrum) should split into four lines [part \textsl{II} in Fig. \ref{fig02}%
(b)], with $\tilde{\omega}_{\circlearrowright }^{\pm }=\tilde{\omega}\pm
g_{\circlearrowright }$ for $\circlearrowright $-polarized and $\tilde{\omega%
}_{\circlearrowleft }^{\pm }=\tilde{\omega}\pm g_{\circlearrowleft }$ for $%
\circlearrowleft $-polarized phonons. In the quantum Hall state at filling
factor $\nu =2$, the transition $-,1\Rightarrow 0$ becomes successively
blocked and no longer affects the frequency of a $\circlearrowleft $%
-polarized phonon, whereas the transition $0\Rightarrow +,1 $ acquires the
maximum strength, thus, increasing the coupling parameter $%
g_{\circlearrowright }$. This leads to the magnetophonon resonance fine
structure consisting of three peaks, with an even larger splitting between
side lines, as sketched in part \textsl{III} in Fig. \ref{fig02}(b).
Interestingly, this may enable one to directly observe lattice modes with a
definite circular polarization. A further increase of the electron filling
factor reduces the side-line splitting which should completely disappear at $%
\nu =6$, after the transition $0\Rightarrow +,1$ becomes blocked by a
complete filling of the $+,1$ LL [Fig. \ref{fig02}(c)]. The same arguments
hold for p-doped graphene, though in this case the roles of $%
\circlearrowright $- and $\circlearrowleft $-polarized modes are
interchanged.

Magnetophonon resonances with other possible inter-LL transitions $%
n\Rightarrow n+1$ occur at much lower magnetic fields, $B_{n}=B_{0}/(\sqrt{n}%
+\sqrt{n+1})^{2}$. For example, a resonant phonon coupling with the
magnetoexciton $\psi _{\mathcal{A};as}(1)$ is expected to occur at $%
B_{1}\approx 5$T. Its description remains qualitatively similar, though for $%
n>0$ the mode splitting is less pronounced because of the $B$-field
dependence of the coupling constants in Eq. (\ref{eq03}). One finds that $%
g_{\circlearrowright }=g_{\circlearrowleft }$ for $|\nu |<2(2n-1)$. At $\nu
=2(2n-1)$, filling of the $n$-th LL starts changing, which reduces splitting
of the $\circlearrowleft $-polarized mode and gives rise to the four-peak
structure. At $\nu =2(2n+1)$, where the $+,n$ LL becomes completely filled,
splitting of the $\circlearrowleft $-polarized phonon vanishes, thus,
resulting in the three-peak fine structure [part \textsl{III} in Fig. \ref%
{fig02}(b)] that would persist up to $\nu =2(2n+3)$. This is because the
splitting of the $\circlearrowright $-polarized modes remains constant up to
the filling factor $\nu =2(2n+1)$, above which population of the $+,(n+1)$
LL starts to suppress the value of $g_{\circlearrowright }$, until the
latter vanishes at $\nu =2(2n+3)$ [see Fig. \ref{fig02}(d)].

In conclusion, we have predicted a filling-factor dependence of the fine
structure acquired by the in-plane (E$_{2g}$) optical phonon in graphene
when the latter is in resonance with one of the inter-LL transitions in this
material. The effect is expected to be most pronounced when the phonon is
resonantly coupled to the $0\Rightarrow +,1$ and $-,1\Rightarrow 0$
transitions, which requires a magnetic field $B_{0}\simeq 30$T. The
predicted mode splitting may be used to measure directly the strength of the
electron-phonon coupling, and also to distinguish between circularly (left-
and right- hand) polarized lattice modes.

We thank D. Abergel, A. Ferrari, P. Lederer, and A. Pinczuk for useful
discussions. This work was suported by Agence Nationale de la Recherche
Grant ANR-06-NANO-019-03 and EPSRC-Lancaster Portfolio Partnership
EP/C511743. We thank the MPI-PKS workshop `Dynamics and Relaxation in
Complex Quantum and Classical Systems and Nanostructures' and the Kavli
Institute for Theoretical Physics, UCSB (NSF PHY99-07949) for hospitality.

\newpage

\section*{Erratum}
In the previous version (v3) of this Letter, we have underestimated
the numerical value of the mode splitting of the magnetophonon
resonance [see paragraph after Eq. (5)] by a factor of $2$ (the text
above takes into account the corrected parameters). This is a result
of two mistakes. First, there is a factor of $\sqrt{2}$, which finds
its origin in an erroneous normalization of the circular polarized
phonons. They should indeed be defined as $u_{\rpol}=(u_x+i
u_y)/\sqrt{2}$ and $u_{\lpol}=(u_x-i u_y)/\sqrt{2}$ [and not as
$u_{\rpol}=u_x+i u_y$ and $u_{\lpol}=u_x-i u_y$ as incorrectly
assumed on page 1, second column], such that the associated phonon
operators $b_{\Amath}$ obey the usual commutation relations
$[b_{\Amath},b_{\Amath'}^{\dagger}]=\delta_{\Amath,\Amath'}$, with
$\Amath=\rpol,\lpol$. This yields a factor of $\sqrt{2}$ in the
definition of the effective coupling constants [Eq. (3)], which read
in the corrected form
\begin{eqnarray}\nonumber
g_{\circlearrowleft }(n)=g\sqrt{(1+\delta _{n,0})\gamma }\sqrt{%
\bar{\nu}_{-,(n+1)}-\bar{\nu}_{+,n}}\ , \\
\nonumber
g_{\circlearrowright }(n)=g\sqrt{(1+\delta _{n,0})\gamma }\sqrt{%
\bar{\nu}_{-,n}-\bar{\nu}_{+,(n+1)}}\ .
\end{eqnarray}
As a consequence, the zero-field dimensionless coupling constant
$\lambda$ [defined in the first column page 3 of our Letter] is
multiplied by a factor of 2 and becomes $\lambda =(2/\sqrt{3}
\pi)(g/t)^{2}$.

Second, we also underestimated the numerical value of the
electron-phonon coupling constant $g$ by a factor of $\sqrt{2}$.
Indeed, $g$ defined in our work [see Eq. (1)] is related to $\langle
g_\Gamma^2\rangle_F\simeq 0.0405$~eV$^2$ computed by Piscanec
\emph{et al.} \cite{piscanecbis} as $g=\sqrt{2\langle
g_\Gamma^2\rangle_F}\simeq 0.28$ eV and not as $g=\sqrt{\langle
g_\Gamma^2\rangle_F}\simeq 0.2$ eV as incorrectly assumed in our
Letter. In addition, there is a substantial uncertainty in the
precise value of the constant $g$. In a tight-binding model, the
latter may be related to the derivative of the hopping amplitude $t$
as a function of the carbon-carbon distance $a$ as $g=(-dt/da)\times
3/(2\sqrt{M\omega})$ \cite{andobis}. Harrison's phenomenological law
$t\propto 1/a^2$ then implies that $g\simeq 0.26$~eV. Experiments in
graphene \cite{pisanabis} and \cite{yanbis} in zero magnetic field
give for the dimensionless coupling constant $\lambda$ the values
$4.4\times10^{-3}$ and $5.3\times10^{-3}$ respectively. This
determines $g$ in between $0.3$ eV and $0.36$ eV, where we take into
account that the value of $t$ lies between $2.7$ and $3$ eV. In the
end, we have to take $g$ in the range between $0.26$ and $0.36$ eV
[instead of $g\simeq 0.2$~eV] and therefore the dimensionless
coupling constant becomes $\lambda\simeq (2.8 \textrm{ to } 5.3)
\times 10^{-3}$ [instead of $\lambda \simeq 10^{-3}$].

As a result of the two factors of $\sqrt{2}$, the numerical estimate
for the mode splitting $2g_{\Amath}$ at $\nu=0$ and $B\simeq 30$ T
[at the discussed resonance $-,1\Rightarrow 0$ and $0\Rightarrow
+,1$, see second column of page 3] becomes $2g_{\Amath}\sim 15$ meV
($\sim 120$ cm$^{-1}$), for $g\simeq 0.26$ eV and $2g_{\Amath}\sim
20$ meV ($\sim 160$ cm$^{-1}$) for $g\simeq 0.36$ eV [instead of
$2g_{\Amath}\sim 8$ meV]. The effect is therefore twice larger than
initially predicted. The conclusions of our work remain unaltered.

\vspace{0.5cm} We would like to thank C. Faugeras and M. Potemski
for having drawn our attention on the underestimated value of the
mode splitting. See also their recent preprint where they measure
the magnetophonon resonance \cite{faugerasbis}.


\begin{thebibliography}{99}
\bibitem{symmetry} The electron-phonon coupling is off-diagonal because a
lattice distortion affects the bond length and thus the nearest-neighbor
hopping between the two different sublattices \cite{castroneto,ando}.
\end{thebibliography}
\end{document}